\documentclass[aps,showpacs,floatfix,superscriptaddress,pre,11pt]{revtex4-1}

\usepackage{amsmath,amssymb}
\usepackage{verbatim,graphicx}

\usepackage{color}
\usepackage[all]{xy}
\newcommand{\Tr}{\rm Tr}
\newcommand{\eref}[1]{(\ref{#1})}
\newcommand{\nn}{\nonumber}
\newcommand{\be}{\begin{eqnarray}}
\newcommand{\ee}{\end{eqnarray}}

\newcommand{\bmat}{\left ( \begin{array}{cc} }
\newcommand{\emat}{\end{array} \right ) }

\def\Tr{\textrm{Tr}}

\newcommand{\beq}{\begin{equation}}
\newcommand{\beqs}{\begin{equation*}}
\newcommand{\eeq}{\end{equation}}
\newcommand{\eeqs}{\end{equation*}}

\allowdisplaybreaks

\begin{document}

%%%%%%%%%%%%%%%%%%%%%%%%%%%%%%%%%%%%%%%%%%%%%%%%%%%%%%%%%%%%%%%%
\title{Analytical Spectral Density of the Sachdev-Ye-Kitaev Model at finite $N$}
\author{Antonio M. Garc\'\i a-Garc\'\i a}
\affiliation{TCM Group, Cavendish Laboratory, University of Cambridge, JJ Thomson Avenue, Cambridge, CB3 0HE, UK}
\email{amg73@cam.ac.uk}
\author{Jacobus J. M. Verbaarschot}
\affiliation{Department of Physics and Astronomy, Stony Brook University, Stony Brook, New York 11794, USA}
\email{jacobus.verbaarschot@stonybrook.edu}
\begin{abstract}
  We show analytically that the spectral density of the $q$-body Sachdeev-Ye-Kitaev model 
  agrees with that of Q-Hermite polynomials with Q a non-trivial function of $q \ge 2$ and the number of Majorana fermions $N \gg 1$. Numerical results, obtained by exact diagonalization, are in excellent agreement
  with the analytical spectral density even for relatively small $N \sim 8$.
  For $N \gg 1$ and not close to the edge of the spectrum,
  we find the macroscopic spectral density simplifies to  $\rho(E) \sim  \exp[2\arcsin^2(E/E_0)/\log \eta]$,  where $\eta$ is the suppression factor of the contribution of intersecting Wick contractions relative to nested contractions. This spectral density reproduces the known result for the free energy in the large $q$ and $N$ limit.
  In the infrared region, where the Sachdeev-Ye-Kitaev model is believed to have a gravity-dual, the spectral density is given by $\rho(E) \sim \sinh[2\pi \sqrt 2 \sqrt{(1-E/E_0)/(-\log \eta)}]$. It  therefore has a square-root edge,
  as in random matrix ensembles, followed by an exponential growth, a distinctive feature of black holes and also of low energy nuclear excitations. Results for level-statistics
  in this region confirm the agreement with random matrix theory. Physically this is a signature that, for sufficiently long times, the SYK model and its gravity dual evolve to a fully ergodic state whose dynamics only depends on the global symmetry of the system. Our results strongly suggest that random matrix correlations are a universal feature of quantum black holes and that the SYK model, combined with holography, may be relevant to model certain aspects of the nuclear dynamics.

%Quantum mechanics produces entanglement but at the same time limits its growth rate.  
\end{abstract}

%\preprint{DAMTP-????}t

\maketitle
\section{Introduction}
Majorana fermions in zero spatial dimensions with $q-$body infinite-range random interactions in Fock space, commonly termed  Sachdev-Ye-Kitaev (SYK) models \cite{kitaev2015,maldacena2016,polchinski2016,engels2016,almheiri2015,magan2016, danshita2016,bagrets2016,sachdev2015,Gross:2016kjj,jensen2016}, are attracting a great deal  of attention as one of the simplest strongly interacting
system
 with a gravity dual \cite{maldacena1998}. Previously, a closely related model with Majorana fermions replaced by Dirac fermions at finite chemical potential was intensively investigated in nuclear physics \cite{bohigas1971,bohigas1971a,french1970,french1971,kota2001,benet2003} and later in the study of spin-liquids \cite{sachdev1993}.

In the limit of a large number $N$ of Majorana fermions  there is already a good understanding of many features of the model including thermodynamic properties \cite{kitaev2015,maldacena2016,jevicki2016}, correlation functions \cite{maldacena2016,bagrets2016,jevicki2016}, generalizations to non-random coupling \cite{witten2016}, higher spatial dimensions and different flavors of Majorana fermions \cite{Gross:2016kjj}. All evidence points to a gravity-dual interpretation \cite{maldacena1998} of the model in the low-temperature strong-coupling limit. More specifically, it is believed that, in this limit, the gravity dual of the SYK model is related to an Anti-deSitter (AdS) background in two bulk dimensions AdS$_2$ \cite{almheiri2015,jensen2016,maldacena2016a} which 
likely describes the low-energy sector of a string-theory dual to a gauge theory in higher dimensions. Related recent work can be found in Refs.
\cite{Berkooz:2016cvq,Fu:2016vas,Klebanov:2016xxf,Nishinaka:2016nxg,Peng:2016mxj,Liu:2016rdi,Turiaci:2017zwd,magan2016a,Kolovsky:2016irf,banerjee2016,krishnan2016,kyono2017}.

One of the main appeals of the SYK model is the possibility to study explicitly finite $N$ effects which are holographically dual to quantum-gravity corrections \cite{kitaev2015,maldacena1998}. 
Indeed, evidence for the existence of a SYK gravity dual is not restricted to large $N$ features such as a finite entropy at zero temperature or a finite specific heat coefficient but also includes properties controlled by $1/N$ effects such as the exponential growth of the spectral density \cite{maldacena2016,cotler2016}, the pattern of conformal symmetry breaking or, for intermediate times of the order of the Ehrenfest time, the universal exponential growth  of certain out-of-time-ordered correlators \cite{maldacena2015,kitaev2015,maldacena2016}. The latter is also a well known feature \cite{larkin1969} of quantum chaos, namely, quantum features of classically chaotic systems. 

Exponential growth of the spectral density together with random matrix correlations of the eigenvalues
is a feature that is also well-known in nuclear physics (see \cite{vonEgidy:1986szw,Haq:1982soh}), in particular for compound nuclei.
These are excited nuclei where the energy of the incoming channel has been distributed over all
nucleons. Because the  dynamics is chaotic all information on the formation of the compound nucleus
is lost, and the quantum state is determined by the total energy and the exact quantum numbers.
In this sense, a compound nucleus has no hair. However, it has ``quantum hair'' in the form
of resonances  which have been measured experimentally \cite{Garg:1964zz}. It turns out that fluctuations
of the compound nucleus cross-section obtained from these experiments agree well with random matrix theory predictions
\cite{Verbaarschot:1985jn}. This implies that the $S$-matrix distribution is determined by causality or analyticity, ergodicity and the maximization 
of the information entropy \cite{mexico}.

Interestingly, qualitatively similar features have recently been found \cite{you2016,garcia2016,cotler2016} for the  SYK model. More specifically, the quantum chaotic nature of the model has been confirmed by showing that for long times scales, of the order of the Heisenberg time, level statistics are well described by random matrix theory \cite{dyson1962a,guhr1998}.
The relation of this finding with features of the gravity dual has yet to be explored as the analysis of spectral correlations carried out in these papers concerns the bulk of the spectrum and not the infrared tail related to the physics of the gravity dual. Moreover the exponential growth of the SYK spectral density, a strong indication of the existence of a gravity dual, is based \cite{kitaev2015,maldacena2016} on a perturbative $1/N$ calculation that may be spoiled by non-perturbative effects.

Here we address these two problems simultaneously. 
We compute analytically the spectral density of the $q-$body SYK model, for any $q$, by 
explicit evaluation of the moments for a large number of fermions.
The combinatorial factors are evaluated explicitly 
by using the Riordan-Touchard formula \cite{touchard1952,riordan1975,flajolet2000}, derived
originally in the theory of cords diagrams. We find that the moments of
the density are equal to those of Q-Hermite polynomials with $Q(N,q)$ a non-trivial function of $N$ and $q$ that we compute explicitly. Agreement with exact numerical results for $N \le 34$ is
excellent in spite of the $N \gg 1$ approximation involved in the analytical calculation. Our calculation follows the steps outlined in Ref. \cite{erdos2014} for a 
closely related spin-chain model
but we keep $q \ge 2$ fixed and $N \gg 1$  rather than considering the scaling limit $N\to \infty $ with $ q^2/N$ fixed. 
In the infrared limit, the spectral density has a square root singularity, as in random matrix
theory. Indeed a detailed analysis of level statistics in this spectral region confirms
excellent agreement with random matrix theory predictions. This suggests that, for sufficiently long times, a quantum black hole reaches a fully ergodic and universal state which only depends on global symmetries of the system. 

Finally we note that the particular case $q \propto \sqrt{N}$ fix and $N \to \infty$  was recently studied  \cite{cotler2016} for the SYK
model where the techniques of Ref. \cite{erdos2014} were also employed to compute the infrared limit of the spectral density.  
By contrast, our results for the spectral density, which agree with those of \cite{cotler2016}, are derived without explicitly taking this double scaling
limit. Therefore they can be applied
to the physically most relevant case $q = 4$ which we also study numerically by exact diagonalization in order to compare with the analytical predictions.

Next we introduce the model, compute analytically the spectral density and compare it with numerical results. We close with concluding remarks and a discussion of our results.

 \section{Model and calculation of the spectral density}
We study $N$ strongly interacting Majorana fermions, introduced in Ref.\cite{kitaev2015}, with infinite range $q$-body interactions. For $q=4$, the Hamiltonian is given by,

\begin{equation}\label{hami}
H \, = \, \frac{1}{4!} \sum_{i,j,k,l=1}^N J_{ijkl} \, \chi_i \, \chi_j \, \chi_k \, \chi_l \, ,
\end{equation}
where $\chi_i$ are Majorana fermions that verify 
\begin{eqnarray}
\{ \chi_i, \chi_j \} = \delta_{ij},\label{clif}
\end{eqnarray}
we note that this is the same algebra as Dirac $\gamma$ matrices which will facilitate the analytical evaluation of the moments. For that reason we will use in many instances the notation $\gamma$ to refer to the fields $\chi$. 

The coupling  $J_{ijkl}$ is a Gaussian random variable 
with probability distribution,
\begin{equation}
P(J_{ijkl}) \, = \, \sqrt{\frac{N^{q-1}}{2(q-1)! \pi J^2}} \exp\left( - \, \frac{N^{q-1}J_{ijkl}^2}{2(q-1)!J^2} \right) \, ,
\end{equation}
where $J$ sets the scale of the distribution.
%For the sake of simplicity, from now on we set $J = 1$.

The average spectral density can be evaluated from the moment generating function
 \be
 \label{rhog}
 \rho(E) &=&
 %\frac 1{N_e} \frac 1{2\pi} \int_{-\infty}^\infty  \left \langle \Tr e^{i(H-E) t} \right \rangle  \nn\\
% &=& 
 \frac 1{2\pi} \int_{-\infty}^\infty e^{- iE t} \left \langle \Tr e^{iH t} \right \rangle.  
 \ee
 %with $N_e$ the number of eigenvalues of $H$. 
 Since the ensemble is invariant
 under ${J} \to -{J}$ we have that $\rho(-E) = \rho(E)$ so that
 the odd moments vanish.
 The moment generating function, given by
 \be
 \left \langle \Tr e^{iH t} \right \rangle = \sum_{k=0}^\infty
 \frac{(it)^{2k}}{(2k)!} 
 \left \langle \Tr H^{2k} \right \rangle,
 \ee
and therefore follows from the moments
 \begin{eqnarray}
 M_{2p}(d)= \langle \Tr  H^{2p}\rangle. \label{moment}
 \end{eqnarray}   
 If we use  the shorthand notation for the Hamiltonian
 \be
 H =\sum_\alpha J_\alpha \Gamma_\alpha,
 \ee
 where $\Gamma_\alpha$ is the product of four $\gamma$ matrices,
 the moments are 
 \be
 \left \langle  \Tr\left(  \sum_\alpha  J_\alpha \Gamma_\alpha \right)^{2p} \right \rangle.
 \ee
 Since we have a Gaussian distribution, the calculation of the average requires to consider all possible
 Wick contractions. After averaging, the result is given by a product
 of pairs of two factors  $\Gamma_\alpha$. If the factors are adjacent 
 we can use that
 \be
 \Gamma_\alpha^2 =1.
 \ee
 If the factors are not  adjacent we have to commute the factors using that
 \cite{garcia2016}
 \be
 \Gamma_\alpha \Gamma_\beta - (-1)^r  \Gamma_\beta \Gamma_\alpha =0 
 \ee
 where $r$ is the number of $\gamma$ matrices that $\Gamma_\alpha$ and $\Gamma_\beta$ have in common. Generally, this is a difficult task because we have to also keep track of correlations with other factors $\Gamma_\alpha$, but the fourth
 and sixth moments can be evaluated exactly \cite{garcia2016}.
 
 The simplest case is the limit $N\to \infty$ for fixed $p$. To leading
 order in $N$, there are no common $\gamma$ matrices, the $\Gamma_\alpha$
 commute and the moments are simply given by 
 \be
 \langle J_\alpha^2\rangle^p 2^{N/2} (2p-1)!!, 
 \label{semi-moments}
 \ee
 which are the moments of a Gaussian distribution \cite{garcia2016}.
 %\begin{figure}\label{fig3}
 %	\includegraphics[width=13cm]{denN32.pdf}
% 	\caption{Tail of the spectral density for $N=32$. Crosses are numerical results from exact diagonalization of the Hamiltonian %(\ref{hami}) for different disorder realizations so that the total number of eigenvalues is about $10^7$. The solid line is the %analytical prediction Eq.(\ref{rhoir}) that has no free parameters. 
% 		The agreement is very good, the small deviation for very negative energies is a consequence of the $N \gg 1$ approximation %necessary for the derivation of Eq.(\ref{eden}).} 
% \end{figure}
 The next simplest case is when we ignore correlations which was also considered in \cite{cotler2016}.
 Let us consider
 \be
 \Tr  ~\Gamma_\alpha \Gamma_\beta \cdots \Gamma_\alpha \Gamma_\beta,
 \ee
 where the dots denote additional factors $\Gamma$. We keep $\alpha $ fixed
 and consider the contribution from the sum over $\beta$.
 Commuting $\Gamma_\alpha $ and $\Gamma_\beta$ gives a factor
 \be
 \sum_{r=0}^q  (-1)^r {q \choose r} {N-q \choose q-r},
 \label{comb}
 \ee
 where $r$ is the number of common $\chi$ fields which, as was mentioned previously, are equivalent to Dirac $\gamma$ matrices. Choosing them out of
 the $q$ $\gamma$ matrices of $\Gamma_\alpha$ gives a factor ${q \choose r}$.
 The remaining $(q-r)$ $\gamma$ matrices in $\Gamma_\beta$ still have 
 to be all different from those in $\Gamma_\alpha$. This gives a factor
 ${N-q \choose q-r}$ resulting in the combinatorial factor of Eq. \eref{comb}.
 If $\Gamma_\alpha $ and $\Gamma_\beta$ were commuting the sum over $\beta$ would give a factor ${N \choose q} $.  Therefore, the suppression
 factor is given by
 \be
 \eta_{N,q} = {N \choose q}^{-1}  \sum_{r=0}^q  (-1)^r {q \choose r} {N-q \choose q-r}.
 \label{suppress}
 \ee
 
 The contractions contributing to the $2p$-th moment can be characterized according to the number of crossings $\alpha_p$. If there are $\alpha_p$ crossings the
 diagram is suppressed by a factor
 $
 \eta_{N,q}^{\alpha_p}.
 $
 The sum over all crossings is evaluated by means of the Riordan-Touchard
 formula \cite{touchard1952,riordan1975} resulting in the following expression for the moments,
 \be
 \frac{M_{2p}}{M_2^p}=   \sum_{\alpha_p} \eta_{N,q}^{\alpha_p} =
 \frac 1{(1-\eta_{N,q})^p} \sum_{k=-p}^p (-1)^k \eta_{N,q}^{k(k-1)/2}
       {2p\choose p+k}.
       \label{moms}
 \ee
 These are the moments of the spectral density $\rho_{QH}$ corresponding to the
 $Q$-Hermite polynomials with $Q = \eta$ \cite{flajolet2000,erdos2014}.
 Therefore,  there is no need to calculate the Fourier transform of the moment generating function in order to compute the spectral density Eq.\eref{rhog}.
  The final result for the spectral density \cite{erdos2014} of the SYK model Eq. (\ref{hami}) is,
  \be
  \label{eden}
  \rho(E) = \rho_{\rm QH}(E) = c_N\sqrt{1-(E/E_0)^2} \prod_{k=1}^\infty
  \left [1 - 4 \frac {E^2}{E_0^2}
  \left ( \frac 1{2+\eta^k+\eta^{-k}}\right )\right],
  \ee
  where $\eta_{N,q} \equiv \eta$ is the suppression factor defined in Eq.\eref{suppress}, $c_N$ is a normalization constant determined by imposing that the total number of states is $2^{N/2}$, and 
  \be \label{e0edge}
  E_0^2 = \frac {4 \sigma^2}{1-\eta},
  \ee
  is the average value of the square of the ground
state energy per particle, i.e. the ground state energy is $N E_0$,
 with the variance  $\sigma$ \cite{garcia2016} given by,
 \be
 \sigma^2 = {N\choose q} \frac {J^2 (q-1)!} {2^qN^{q-1}}.
 \ee
   We note that the product in Eq. \eref{eden} can also be expressed in terms
 of a $q$-Pochhammer symbol. 
 
 A natural question to ask is the precise requirements for the validity of Eq.
 (\ref{moms}).
 While we cannot give a rigorous proof, we believe that, for $p \ll N $ fixed, Eq. (\ref{moms}) agrees with the exact moments including $1/N$ corrections. Still we can get in principle large corrections to the analytical prediction when the order of the moments
 becomes of order $N$. In general the high moments have a strong impact
 on extreme 
 eigenvalues which control the zero temperature entropy and specific heat coefficient.
 We can explicitly see the validity of Eq.  \eref{moms}  by  analytically
 computing moments to low order.  
 
  For large $N$, the exact result for $M_6/M_2^3$  (see Ref. \cite{garcia2016})
 is very close to the approximate result \eref{moms}, even for $q=2$ and
 $q =4$, with a difference that scales as $1/N^2$, while the fourth moment
 is reproduced identically.
 Exact results for higher
   order moments are not known, but the results below indicate the moments
   \eref{moms} are very close to the exact results.
In Ref. \cite{cotler2016}, instead of using the
  exact suppression factor Eq.~(\ref{suppress}), $\eta$ was approximated by a Poisson
  distribution which is valid in the scaling limit where $q^2/N$ is kept fixed for $N \to \infty$ but not for general 
  $q$.

  For large $N$, at fixed $q$ only the $m=0$ and $m=1$ terms contribute
  to the sum of 
  the  suppression factor Eq.~(\ref{suppress}) resulting in
  \be
  \eta^N
  %&=& 1 - \frac {q^2}N + \frac 12 \frac {q^4}{N^2} - \frac1{3!} \frac {q^6}{N^3} +\cdots \nn\\
  &\sim& e^{-2q^2},
  \ee
  where we have used that for $N \gg q$ we can make the expansion
  \be
  \frac { \Gamma^2[N-q]}{\Gamma[N]\Gamma[N-2q]} = 1 - \frac {q^2} N + O(1/N^2).
  \ee
  This corresponds to the Poisson distribution used in Ref. \cite{cotler2016}.

  Below we will show, by comparison to exact numerical results, that the above expression for
  the spectral density is close to the exact numerical result for $q = 4$, even for values of $N$
  as low as $N = 8$, where the suppression factor is negative. Before that we work out simplifications of the spectral density Eq.~(\ref{eden})
 valid in the tail and the bulk of the spectrum.

 \subsection{Simple form of the Spectral Density for $N \gg 1$}

 In this section we derive a simple asymptotic form for the spectral density.
 The derivation follows the steps in Ref. \cite{cotler2016}, but we keep $q \ge 2$ fixed and do not
 take the limit $E \to E_0$. This way we obtain an approximate form that
 is valid over the entire spectrum of the Hamiltonian, except very close to the edge, and for any $q$ with the only assumption of $N \gg 1$. For completeness
 we reproduce the steps given in Ref. \cite{cotler2016}.
 
            Writing the product in Eq.\eref{eden} as the logarithm of a sum, we obtain after a Poisson
           resummation
           \be
           \rho_{\rm QH}(E) =c_N \exp \left [\frac 12 \sum_{n+-\infty}^\infty \int dx e^{2\pi i nx }
             \log \left [ 1 -  \frac {E^2}{E_0^2}
           \left ( \frac 1{\cosh^2 x/2 \log \eta }\right )\right]\right ].
             \ee
             The integral over $x$ can be performed analytically resulting in
           \be
           \rho_{\rm QH}(E) = c_N\exp \left [-\frac 12 \sum_{n=-\infty}^\infty
             \frac{1 -\cosh[\frac{4n\pi}{\log\eta}\arcsin(E/E_0)}
               {n\sinh(2n\pi^2/\log\eta)} \right ]. 
             \ee
             The $n=0$
            term in the sum has to be treated separately as the limit $n\to 0$.
            For $N \to \infty$ we have that $\eta \to 1$ so that for $n\ne 0$,
            we can
            approximate the hyperbolic functions by a single exponent
            leading to
 \be
            \rho_{\rm Bethe} (E)& =& c_N\exp \left [\frac {2\arcsin^2(E/E_0)} { \log \eta}
              +
             \log\left (1 -\exp\left[- \frac{2\pi}{\log\eta}(|\arcsin(E/E_0)|-\frac \pi 2)\right ]\right ) \right ]
              \nn \\
               &=&c_N\exp \left [\frac {2\arcsin^2(E/E_0)} { \log \eta} \right ]
           \left (1 -\exp\left[ - \frac{4\pi}{\log\eta}(|\arcsin(E/E_0)|-\frac \pi 2)\right ]
           \right ).
           \label{rhobethe}
               \ee
               For $N\to \infty$  the second
               factor can be ignored for $|E| < |E_0|$ resulting in a very
               simple
               asymptotic form for the spectral density
\be   \label{rhosim}
            \rho_{\rm asym}(E)
               &=&c_N\exp \left [\frac {2\arcsin^2(E/E_0)} { \log \eta} \right ],
               \ee
               which for finite $N \gg 1$ is an excellent approximation of the spectral density except in the region close to the edge $E_0$. Here a different a
               asymptotic expression can be worked out by simply noticing 
               that for $E\to E_0$, $\arcsin(x) $ is approximated by
               \be
               \arcsin[E/E_0]= \frac \pi 2 - \sqrt 2 \sqrt{1-(E/E_0)}.
               \ee
               Inserting this in Eq. \eref{rhobethe} gives 
\be
               \rho_{\rm edge}(E)
       %        &=&c_N\exp \left [\frac {2(\frac \pi 2 - \sqrt 2 %\sqrt{1-(E/E_0)} } { \log \eta} \right ]
       %     \left (1 -\exp\left[  \frac{4\pi}{\log\eta}\sqrt 2 %\sqrt{1-(E/E_0)}\right ]
       %       \right )\nn\\
              &\approx& c_N\exp \left [ \frac {\pi^2}{2\log\eta}
                - \frac {2\pi\sqrt 2 \sqrt{1-(E/E_0)}}{\log\eta}\right ]
              \left (1 -\exp\left[  \frac{4\pi}{\log\eta}\sqrt 2 \sqrt{1-(E/E_0)}\right ]\right )
              \nn  \\
              &=& 2 c_N\exp \left [ \frac {\pi^2}{2\log\eta}\right ]
              \sinh \left [
                 \frac {2\pi\sqrt 2 \sqrt{1-(E/E_0)}}{-\log\eta}\right ].
           \label{rhocot}
              \ee
              For the limiting case  $q, N \to \infty$ with $q^2/N$ fixed, and still $E \to E_0$, this expression of the spectral density was also obtained in Ref. \cite{cotler2016}.

%                             In Fig. \ref{fig:denana} we show the %$Q$-Hermite form of the spectral density,
%               $ \rho_{\rm QH}(E)$ (black), the approximate form
%               $\rho_{\rm Bethe}(E)$ (red dashed)  and
%               the asymptotic expression, $\rho_{\rm asym}(E)$ (blue %dotted). Results are given for $N=18$, $N=24$, $N=32$ and $N=64$.
%               For $N=32$ the asymptotic form \eref{rhosim}
%               is only barely distinguishable from the exact result in Eq. %\eref{eden}, while
%               for $N=64$ it 
%               can be used all the
%                   way to the edge of the spectrum.
 
               \begin{figure}[t!]
                 \label{compare}
                 \centerline{\includegraphics[width=8cm]{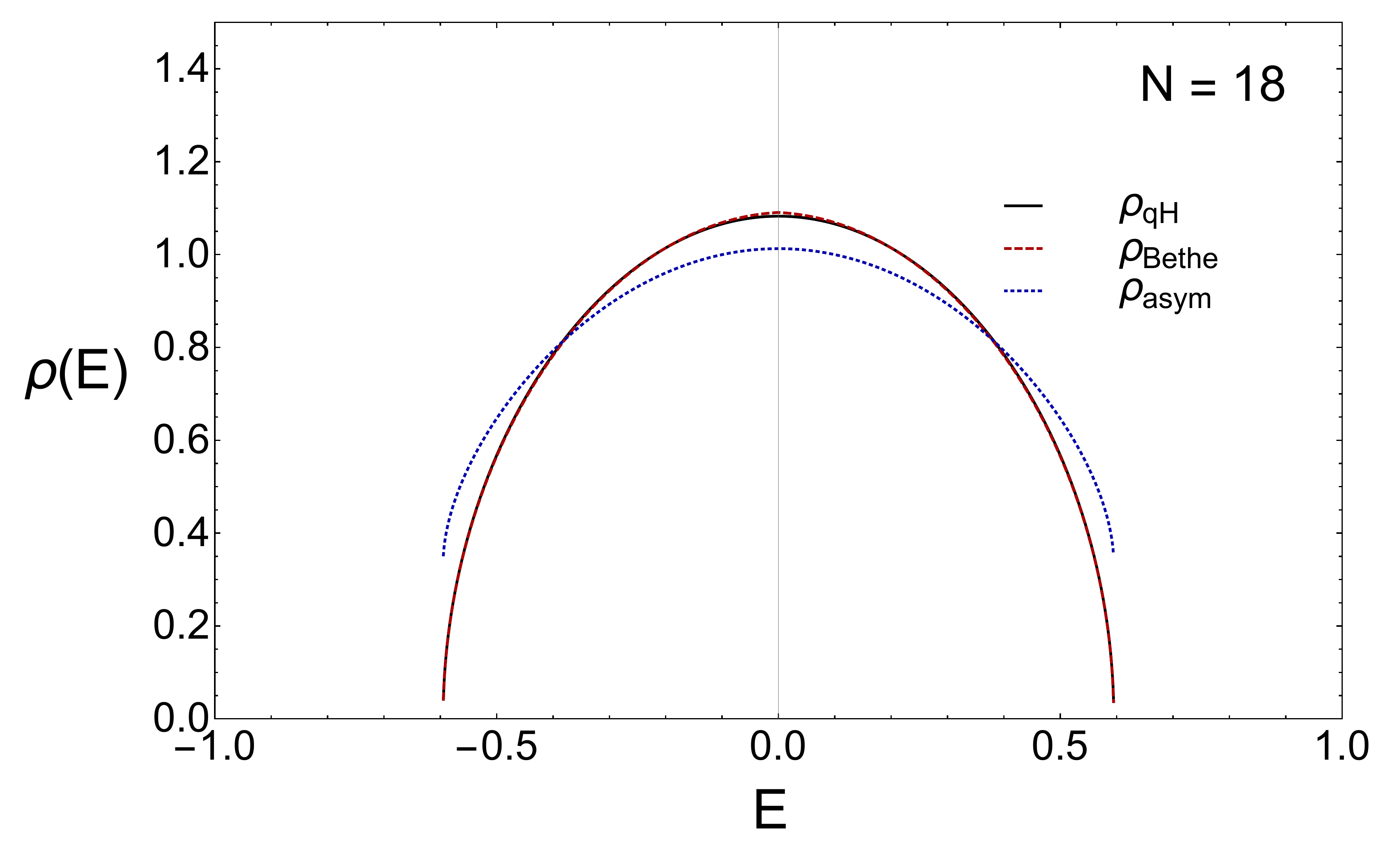}
                   \includegraphics[width=8cm]{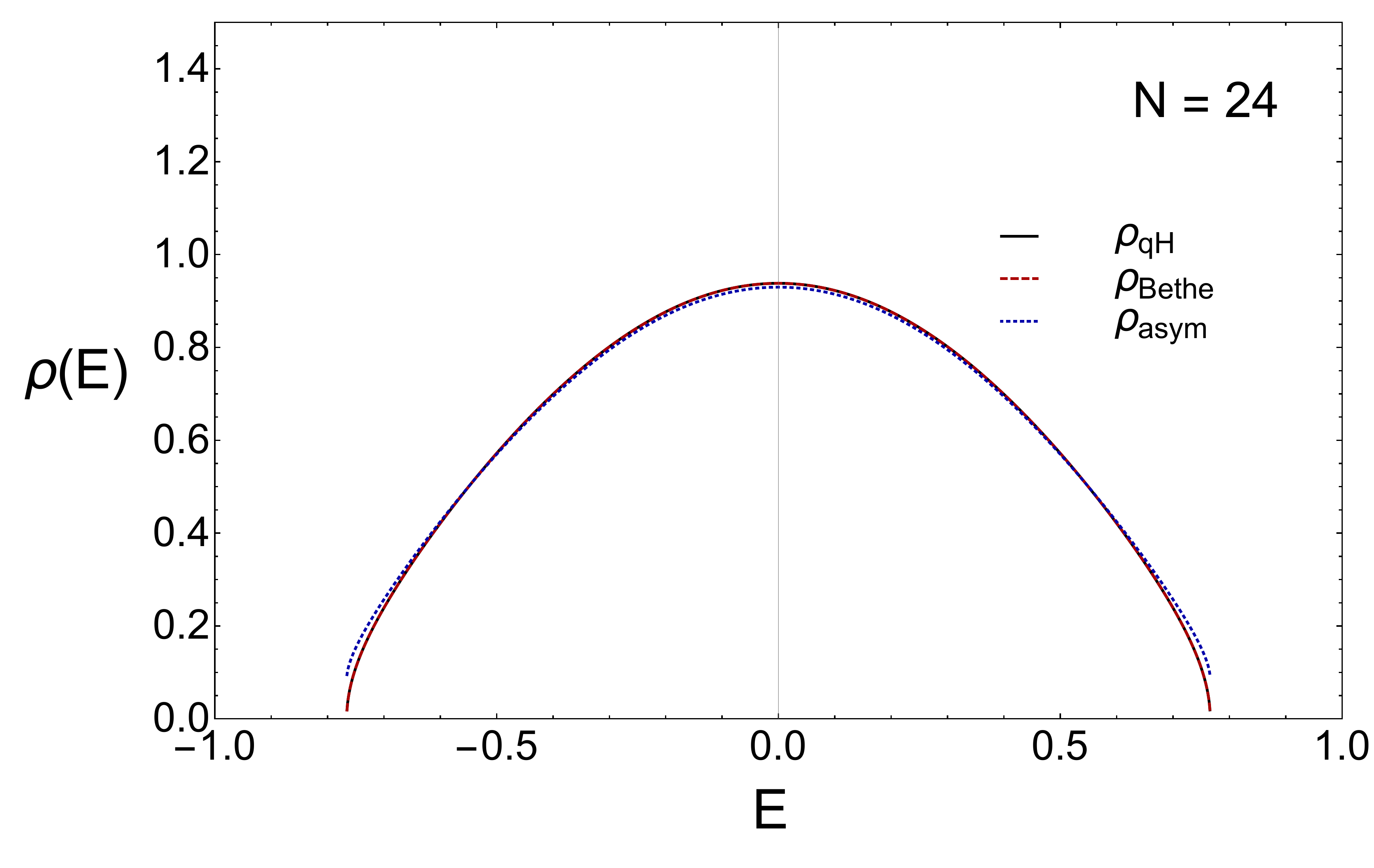}}
                 \centerline{\includegraphics[width=8cm]{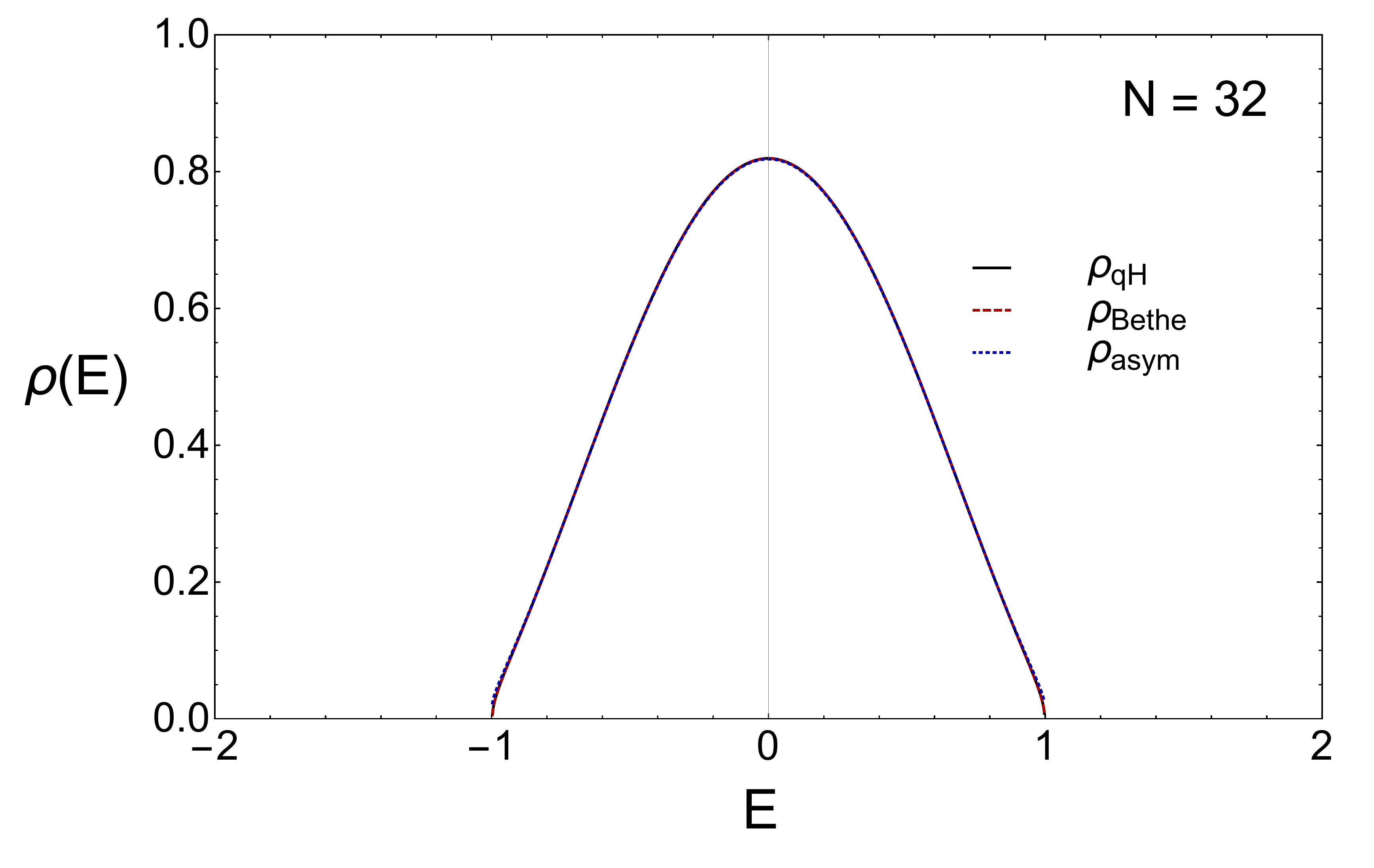}
                   \includegraphics[width=8cm]{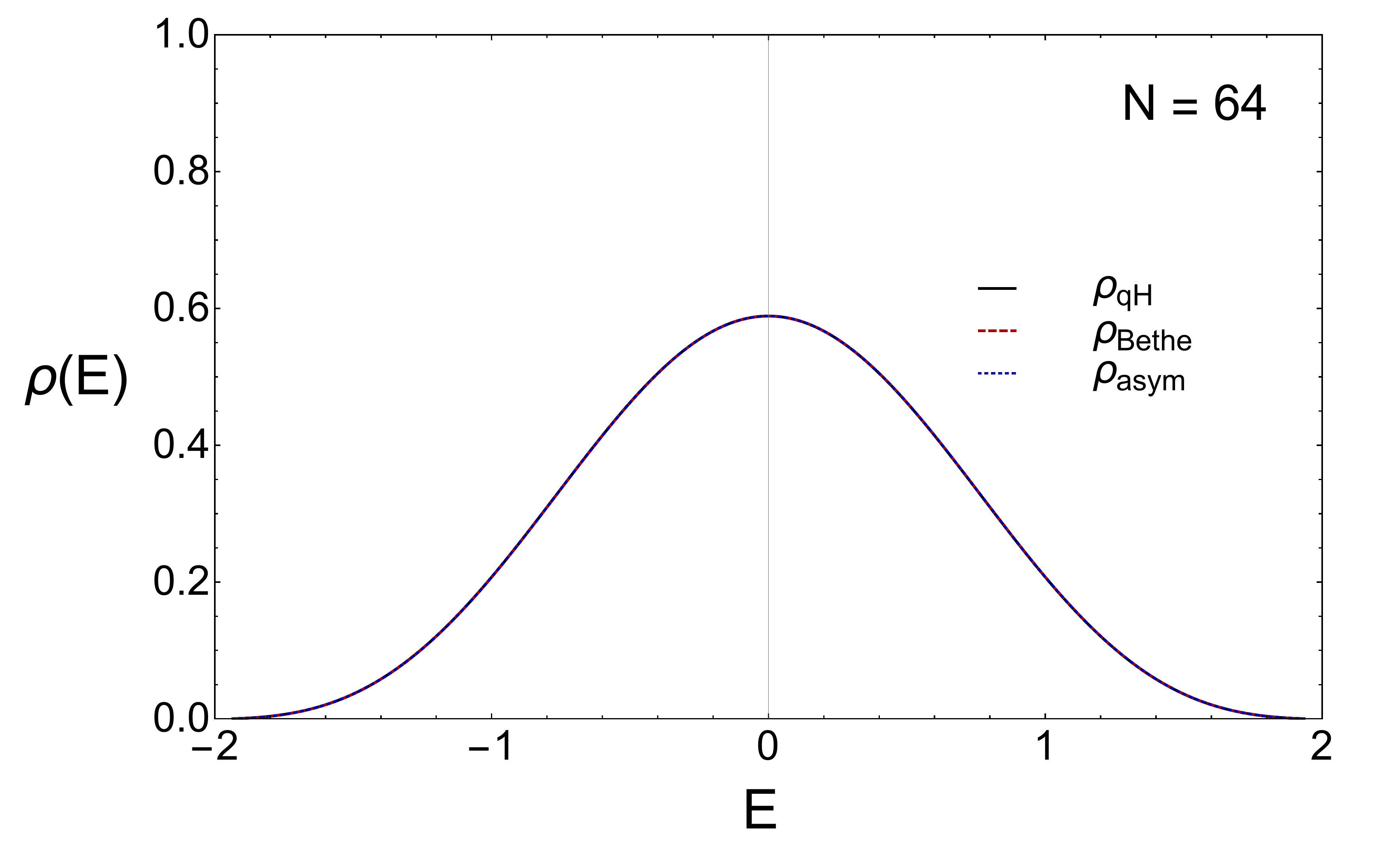}}
                 \caption{In this figure we compare $Q$-Hermite spectral
                   density $\rho_{\rm QH}(E)$  Eq.~(\ref{eden}) (black), of the SYK Hamiltonian to two different asymptotic
                   forms, $\rho_{\rm Bethe}(E)$ Eq.(\ref{rhobethe}) (red dashed)
                   and $\rho_{\rm asym}(E)$ Eq.(\ref{rhosim}) (blue dotted) all
                   normalized to area one.
                   Results
                   are given for $N=18$, $N=24$, $N=32$ and $N=64$.
                   For $N \geq32$ the three curves are barely distinguishable. In all plots the spectral density is normalized to $1$ and $J=2/3$. We note that this is also the value of $J$ in our previous paper \cite{garcia2016}.
                 } \label{fig:denana}
               \end{figure}

We stress this asymptotic form is an expected feature of field theories with a gravity dual as this exponential growth is observed in both systems with conformal symmetry and black-holes. 
The same exponential growth has also been predicted for the low energy excitations of nuclei \cite{bethe1936}.  
 
Having derived the analytical results we now proceed to compare the approximate spectral densities, Eqs.~(\ref{rhobethe}),~(\ref{rhosim}), with the exact $Q$-Hermite form Eq.~(\ref{eden}). Results depicted in Fig. \ref{fig:denana} for different sizes $N$ show that the simple asymptotic expression Eq.~(\ref{rhosim}) agrees reasonably well with the exact result even for comparatively small $N = 18$. Indeed it is barely distinguishable from the exact result Eq. \eref{eden} for $N = 32$ while for $N=64$ it can be used all the way to the edge of the spectrum.

\begin{figure}[t!]
                 \centerline{\includegraphics[width=8cm]{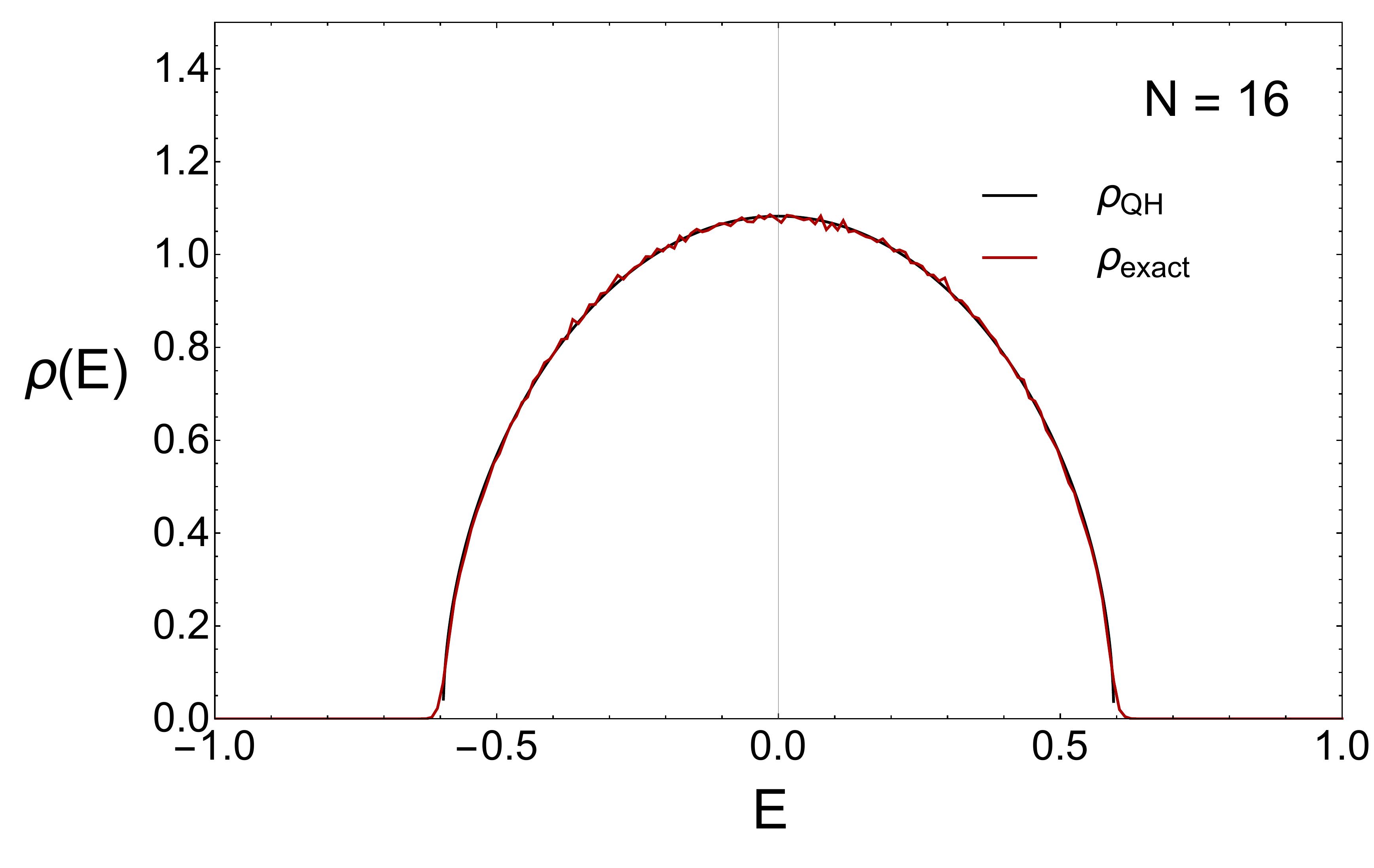}
                   \includegraphics[width=8cm]{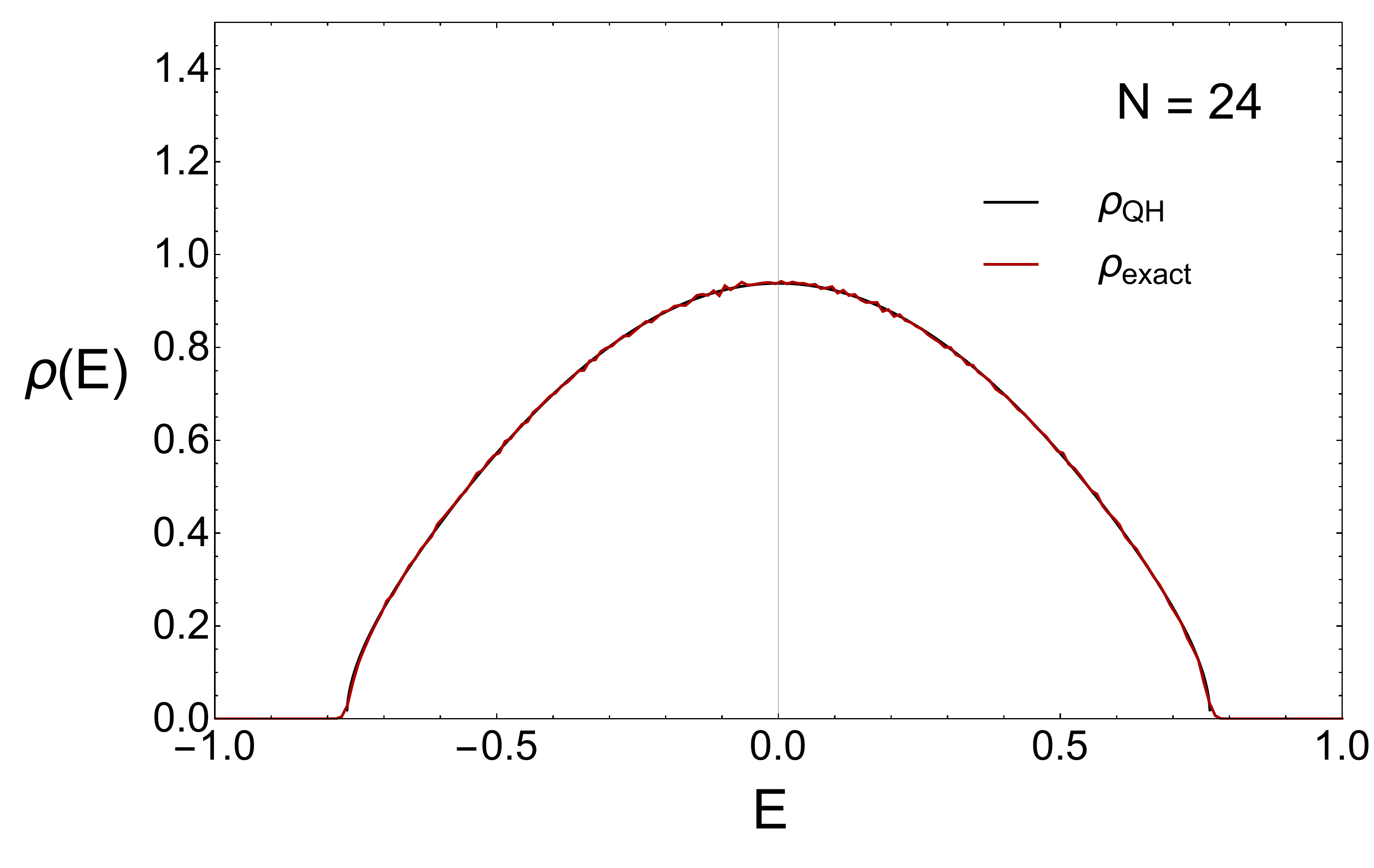}}
                 \centerline{\includegraphics[width=8cm]{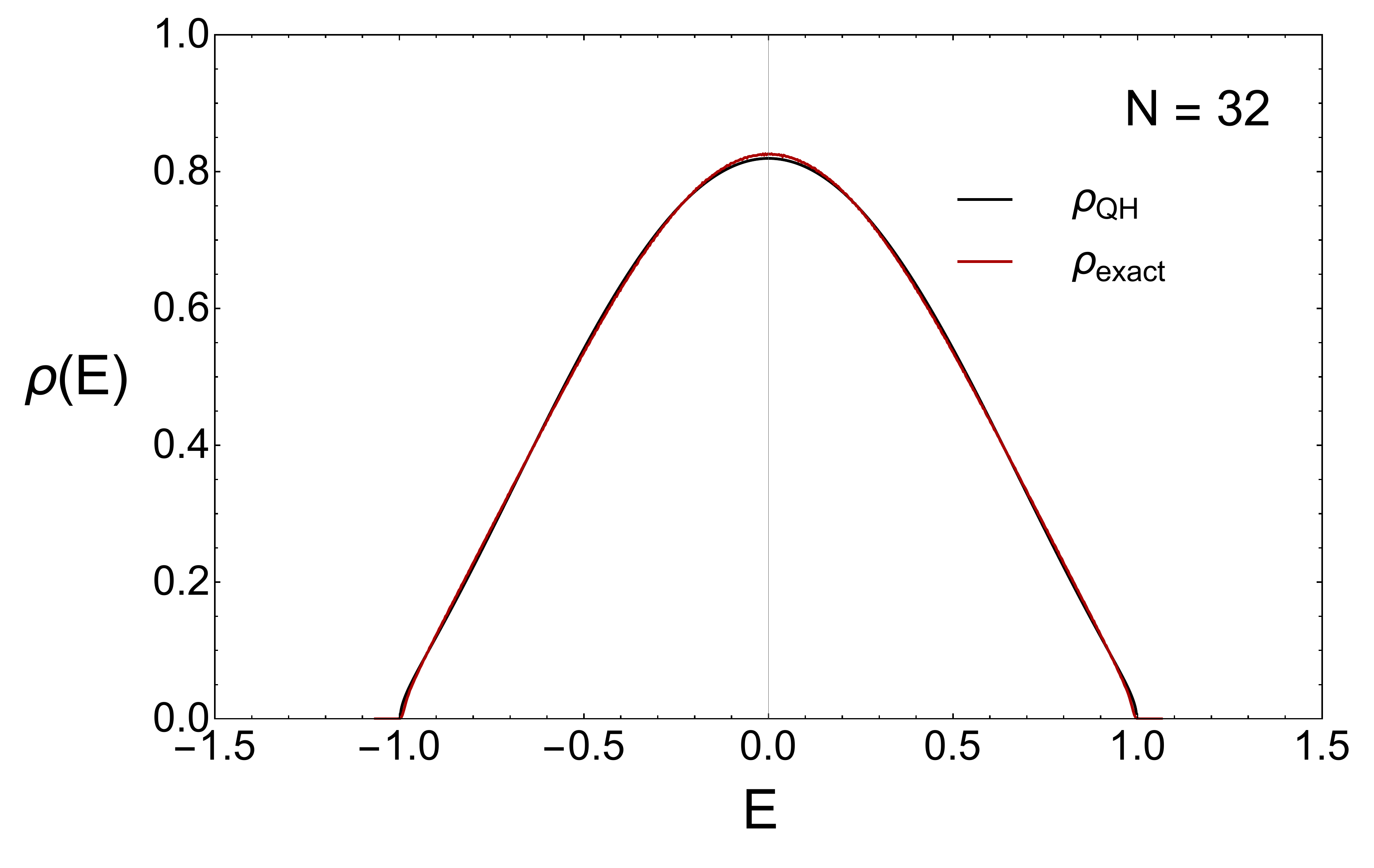}
                   \includegraphics[width=8cm]{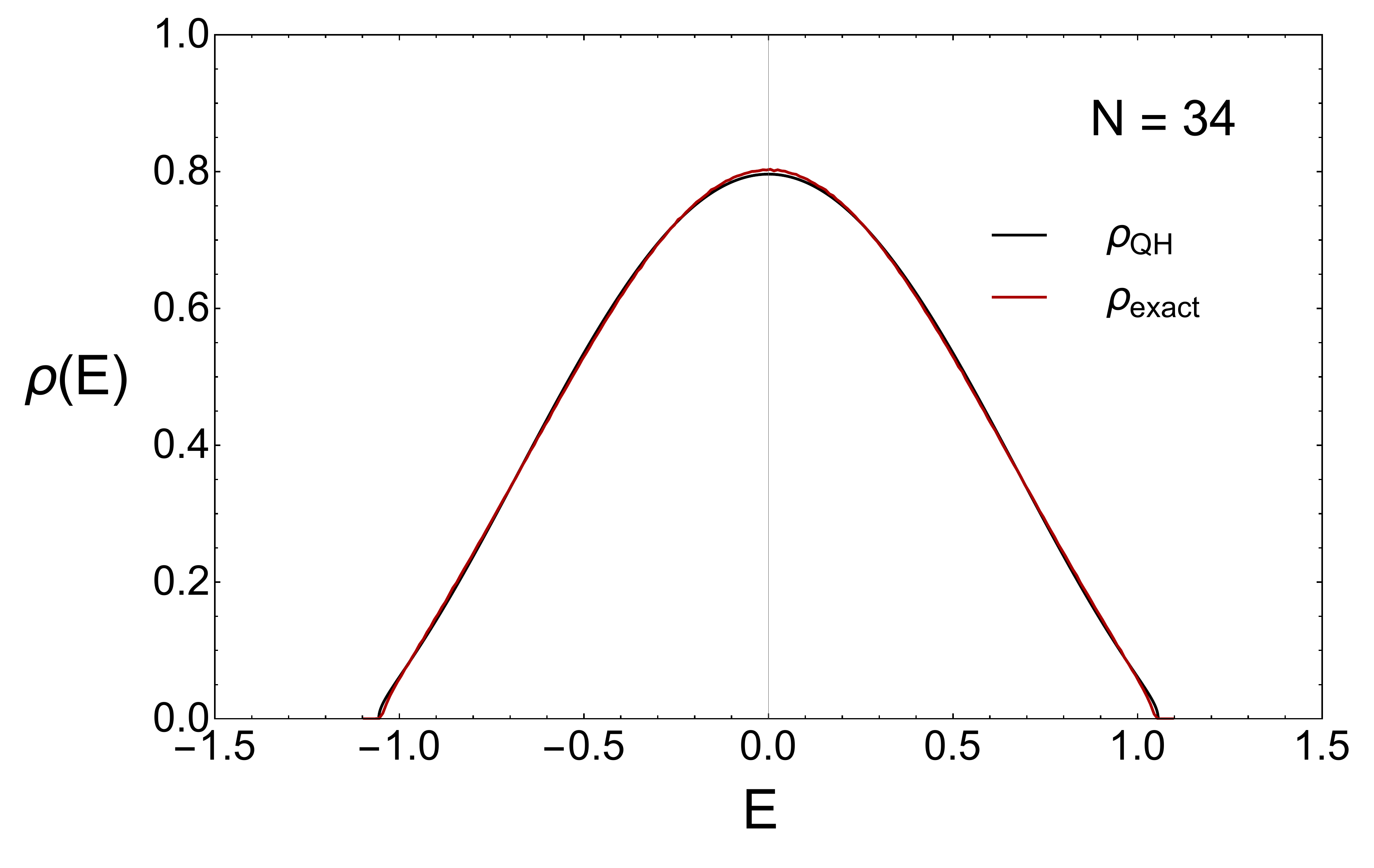}}
                 \caption{Comparison of the numerical spectral density (red) of the SYK Hamiltonian Eq.~(\ref{hami}) for $N=16$, $N=24$, $N=32$ and $N=34$, obtained by exact diagonalization, with the analytical prediction $\rho_{\rm QH}(E)$ Eq.~(\ref{eden}) (black). The agreement is excellent. Even though there is no free parameters the curves are almost indistinguishable. As in the previous figure the spectral density is normalized to 1 and $J = 2/3$. }

                 \label{fig:dencom}
                 \end{figure}

We now proceed to compare these analytical results with numerical results from exact diagonalization of the Hamiltonian Eq.~(\ref{hami}).
By using standard exact diagonalization routines in MATLAB we have obtained the full spectrum of the Hamiltonian Eq.~(\ref{hami}) for many disorder realization so that, for a given size $N \le 34$, the total number of eigenvalues was more than $10^7$.
In Fig. \ref{fig:dencom} we show the exact numerical spectral density (red) and compare it to
the analytical result Eq.~(\ref{eden}) for $N=16$, $N=24$, $N=32$ and $N =  34$. The agreement is excellent.

\begin{figure}[t!]
 	\centerline{
 	\includegraphics[width=7.5cm]{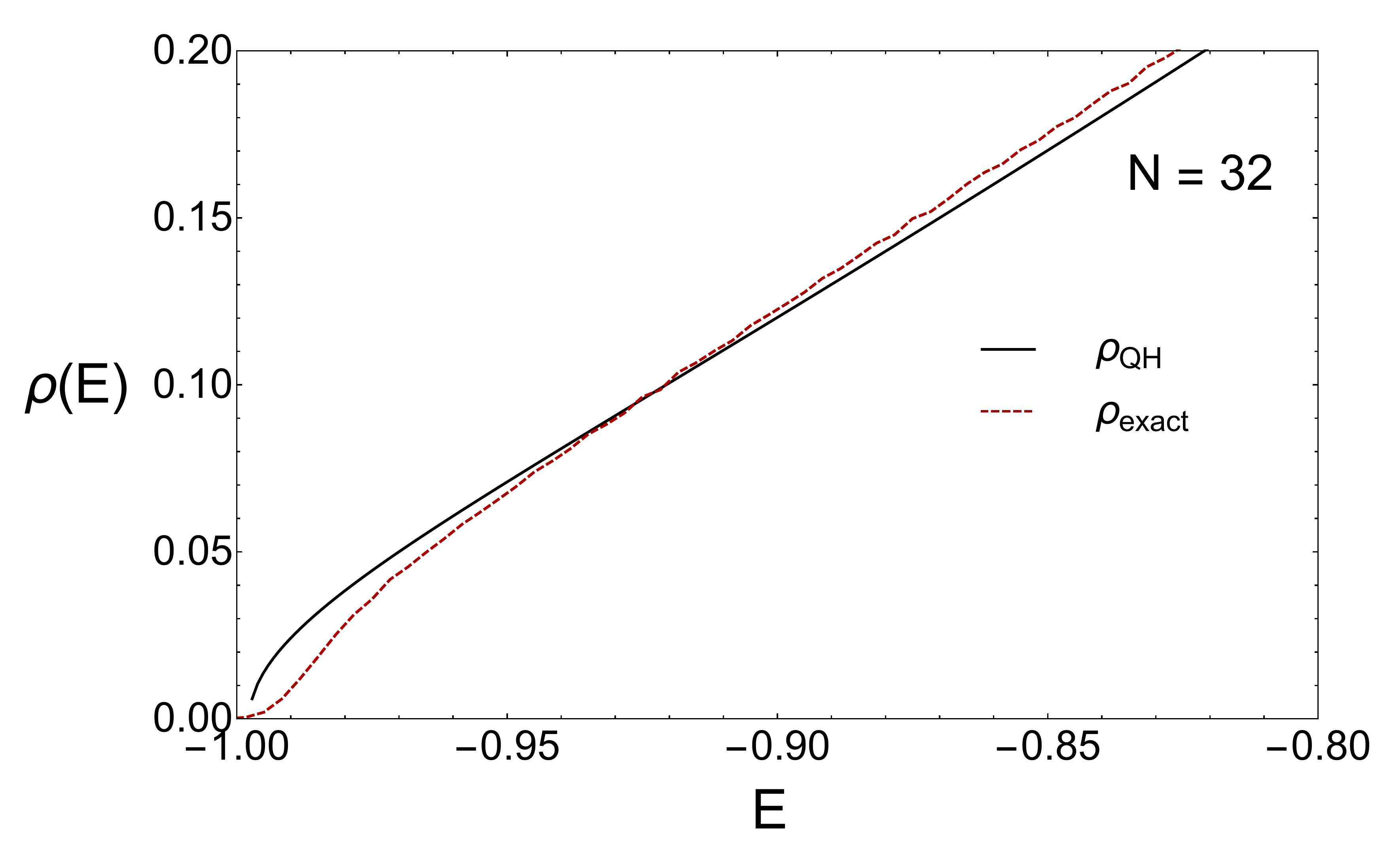}
 	\includegraphics[width=7.5cm]{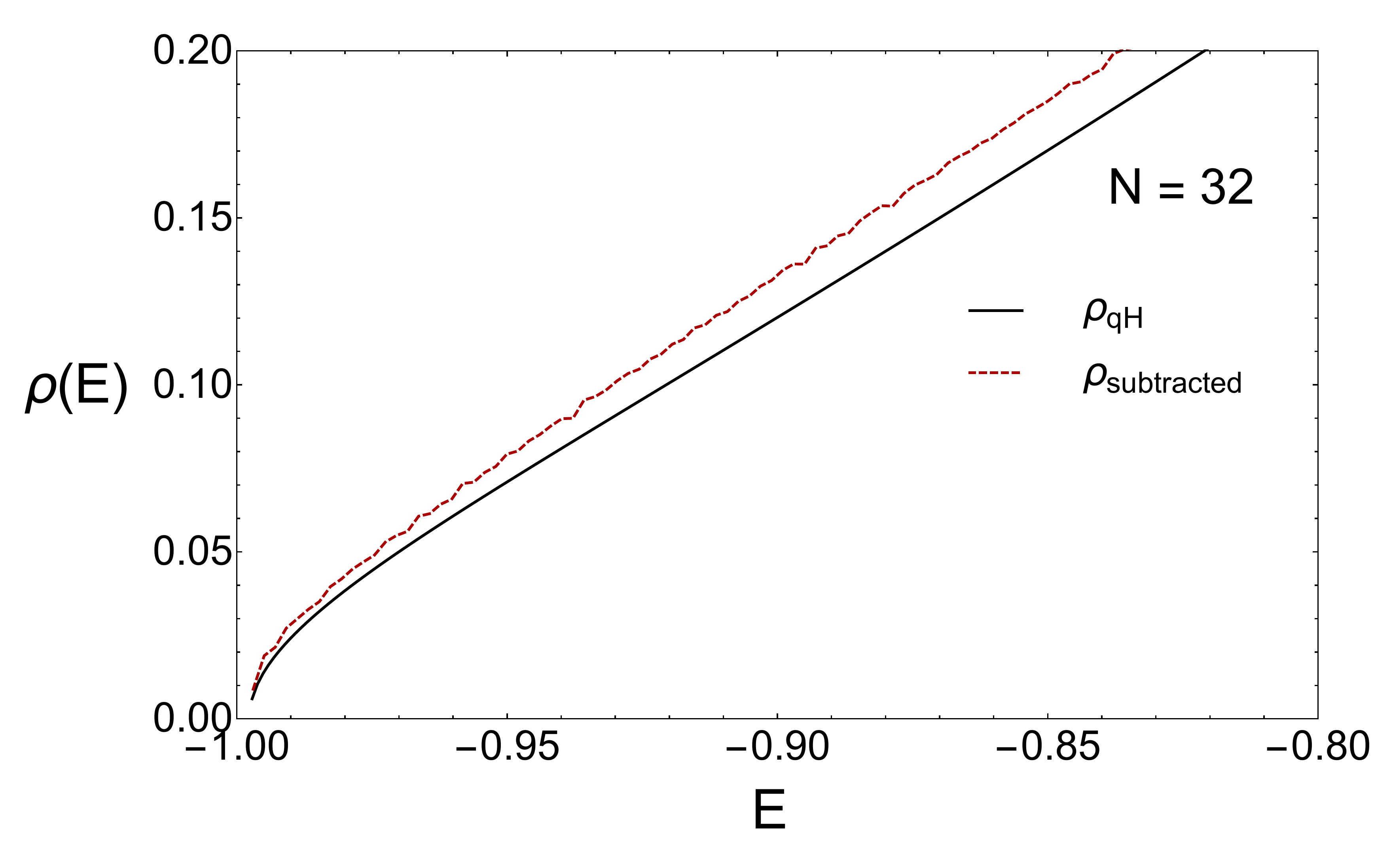}}
        \caption{The tail of the spectral density for $N=32$ and $400$ disorder realizations. In the right figure,
          $E_0-\langle E_0 \rangle$ has been subtracted from all eigenvalues, while
          in the right figure no subtractions have been made.
          The agreement is excellent despite the fact finite $N$ effects, not fully captured in our theoretical analysis, should be stronger in this region. 
          Even without this subtraction the agreement is still very good.
        }
        \label{fig:tail}
\end{figure}

In order to further clarify the extent of the accuracy of the analytical spectral density, we extend the comparison, left plot of Fig.~\ref{fig:tail}, to the deep infrared part of the spectrum where finite size effects are expected to be more relevant.   

 The numerical density is still very close to the analytical prediction but we have found some deviations. For instance the hard edge, predicted analytically, is replaced by a smooth tail. 
 Remarkably, the analytical edge of the spectrum Eq.~(\ref{e0edge}), is still surprisingly close to the
 numerical result. Since not all sub-leading $1/N$ corrections were included in the
 derivation of the spectral density, stronger discrepancies were expected
 for the values of $N$ we work with. It is actually rather unexpected that the analytical result is so close to the numerical calculation.

Still we would like to understand why a tail, and not an edge, is observed in the numerical spectral density.
We shall see in next section that the level statistics of the model in this infrared region are still described by random matrix theory. We note that because of the stiffness of the spectrum, eigenvalues in random matrix
theory fluctuate ``collectively'', which, due to ensemble average, smoothes out the edge of
the spectrum. This is particularly true for the lowest eigenvalue $E_0$, which is a stochastic variable,
while the theoretical prediction Eq.~(\ref{e0edge}) is the  ensemble average. In order for a more accurate comparison one has to either take into account the distribution of $E_0$ or simply remove the fluctuations of $E_0$. We choose the latter. 
 In the right plot of Fig. \ref{fig:tail} we show the spectral
 density relative to the first eigenvalue. To have the same scale on the
 $x$-axes we have added the ensemble average of the first eigenvalue to all
 eigenvalues. This clearly reveals the square root edge of the average spectral density predicted theoretically.

\begin{figure}[t!]
 	\centerline{
 	\includegraphics[width=7.5cm]{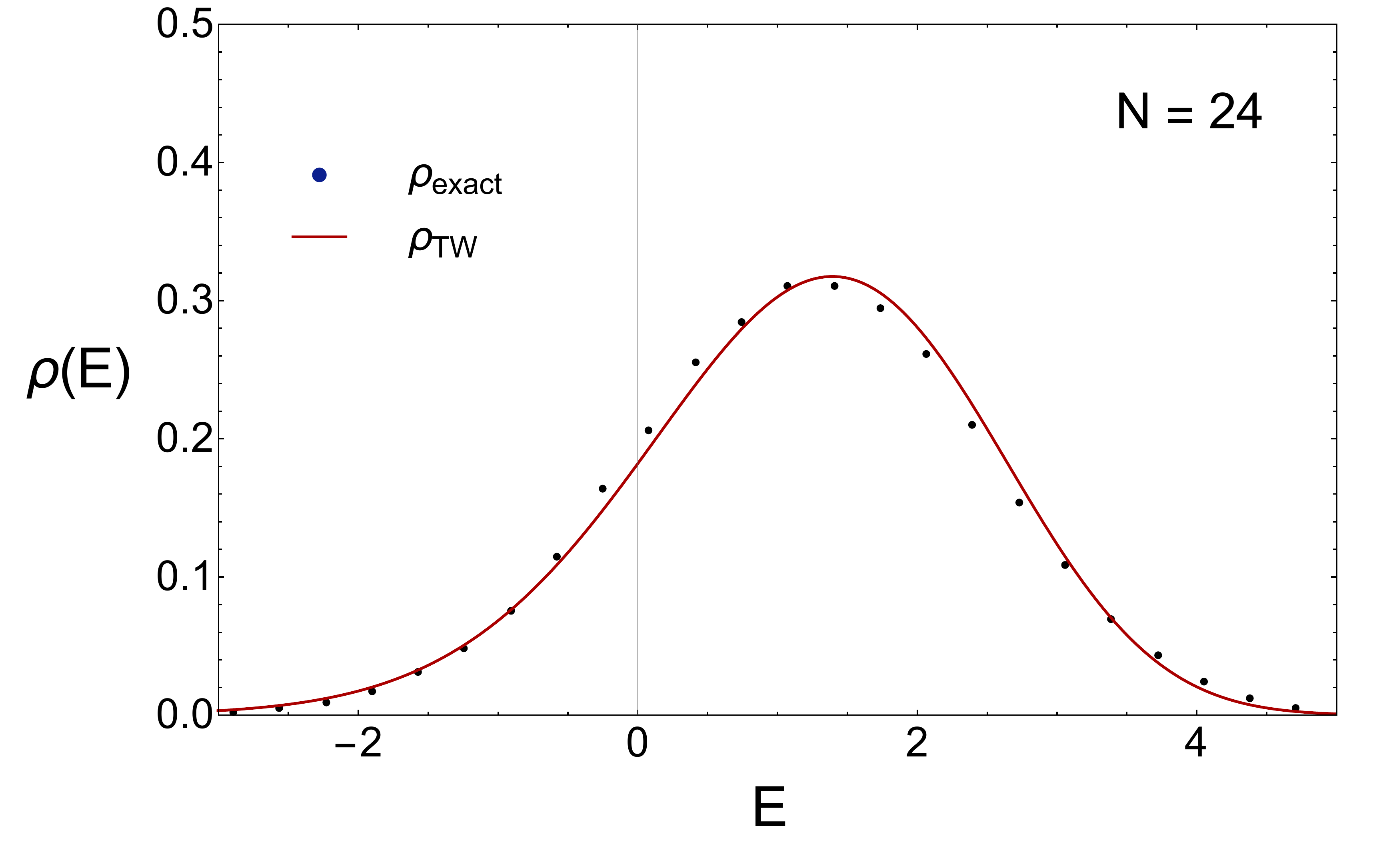}
        	\includegraphics[width=7.5cm]{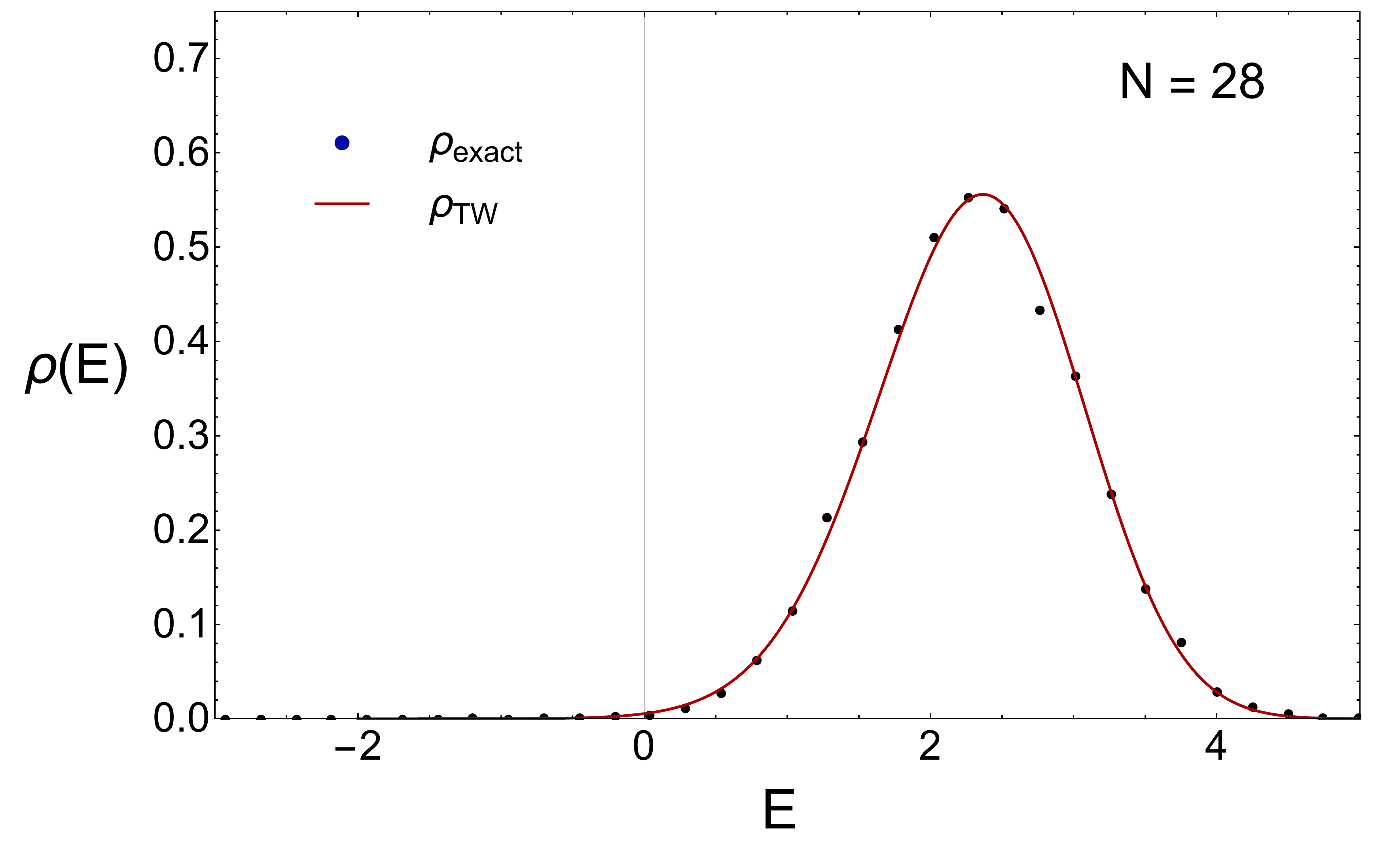}}
        \caption{Distribution
          of the lowest eigenvalue for $N = 24$ (left) and $N=28$ (right) for an
          ensemble of $50,000$ and $15,000$ disorder realizations, respectively,
          compared to
          the random matrix prediction for the Tracy-Widom distribution.
          The numerical data have been shifted and rescaled to reproduce the
          average and variance of the Tracy-Widom distribution.
          The agreement is excellent which confirms that the low energy limit of the SYK model is fully ergodic and well described by random matrix theory.
          } \label{fig:tw}
\end{figure}

 This finding leads us to the prediction that
 the distribution of $E_0$ is the one given by random matrix theory for the distribution of the smallest eigenvalue, namely, the Tracy-Widom distribution \cite{tracy1994}. In Fig.~\ref{fig:tw} we show the distribution of the smallest eigenvalue of the SYK model and compare it
 to the Tracy-Widom distribution of the corresponding random matrix ensemble.
 Results are given for $N=24$ (left), which is in the universality class of the
 Gaussian Orthogonal Ensemble and $N = 28$  (right), which is in the universality class of the Gaussian  Symplectic Ensemble. There are no fitting parameters
 but  the numerical data have been shifted and rescaled to reproduce
 the average and variance of  Tracy-Widom distribution.
 We find good agreement which is another indication that the spectrum
 of the SYK Hamiltonian has a square root edge.

 We now employ the analytical form of the spectral density to study the free energy. We start with the density Eq. \eref{rhosim} which is valid everywhere except in the tail.  
             The partition function in this case is given by
            \be
            Z(\beta) = \int_{-E_0}^{E_0}c_N e^{-\beta E +  \frac {2\arcsin^2(E/E_0)}
            { \log \eta}} .
            \ee
            For $ \log \eta \to 0$, the partition function can be evaluated by a
            saddle point approximation resulting in the free energy
                          \be
              \beta F = \beta \bar E - \frac {2\arcsin^2(\bar E/E_0)} { \log \eta}
              \ee
              where $\bar E$ satisfies the saddle point equation
                          \be
            \beta =\frac 4{ E_0 \log\eta }
            \frac{\arcsin (\bar E/E_0)}{\sqrt{1-(\bar E/E_0)^2}}.
            \ee
            If we define the new variable
                          \be
              \arcsin \frac {\bar E}{E_0} = \frac {\pi v}2,
              \ee
              the saddle point equation can be written as
                \be
                \beta  \mathcal J = \frac{\pi v}{ \cos \frac {\pi v}2}
\label{saddle-m}
                \ee
   with  $\mathcal J  =  (E_0 /2) \log\eta$.                   In terms of these variables, the free energy at the saddle point is given by
\be
\beta F = \frac {2}{\log \eta} \pi v \tan \frac{\pi v}2-\frac{ (\pi v)^2}{2 \log \eta}.
\ee
In the large $N$ limit we have that $\log \eta \to  -2q^2/N$
and this expression together with Eq. \eref{saddle-m} reduces to the result obtained in Ref. \cite{maldacena2016} which, strictly speaking, is only valid in the large $q$ limit. In the low-temperature limit, the fluctuations about
the saddle point gives a factor $1/\beta^{3/2}$ resulting in the low-temperature
limit of the partition function \cite{maldacena2016}
\be 
Z(\beta) \propto \beta^{-3/2}\exp\left [ \beta |E_0| +\frac N2 \log 2
  + \frac {\pi^2} {2\log\eta}    +\frac {2\pi^2}{\beta | E_0| \log^2 \eta } \right ].
\label{zlow}
\ee
We note that the analytical evaluation of the partition function related to the tail of the spectrum Eq.(\ref{rhocot}), that includes $1/N$ corrections, 
 reproduces this result identically.
%Note that the power of the prefactor agrees with the result
%of \cite{maldacena2016} obtained by employing completely different methods. 
%obtained by considering  $1/N$ corrections to the mean-field  limit in the low temperature, strong coupling limit where the SYK model is well approximated by the Schwarzian action.
%We note that the specific heat coefficient in the large $N$ limit has the same $q$ dependence as the analytical result 

In conclusion, the analytical form of the spectral density, which in principle is only valid for sufficiently large $N$ as it only includes $1/N$ corrections, agrees unexpectedly well with exact numerical results. We do not have a clear understanding of the reason behind this suppression of quantum effects but it seems that our analytical results are close to exact. This is especially surprising close to the edge of the spectrum since finite $N$ effects, not fully considered in the theoretical analysis, are expected to be stronger in this region. We can only speculate that in systems with infinite range interactions a mean field approach becomes exact in the large $N$ limit and therefore, for finite $N$, fluctuations may be weaker than in systems with short-range interactions. 

%Finally, our analytical result for
%the spectral density reproduces the result for the free energy obtained
%in Ref. \cite{maldacena2016}.

\section{Applications in nuclear physics and holography}
The SYK and related models have been employed to study different aspects of nuclear physics, condensed matter and, more recently, holographic dualities. We now discuss how the results of the previous section help understand better these systems. We start with holographic dualities. It was previously known \cite{kitaev2015,maldacena2016a} that $1/N$ corrections, combined with the saddle point approximation, lead
to a spectral density that grows exponentially for energies close, but not too close, to the ground state energy. This is considered to be a distinctive feature of quantum black holes
in the semiclassical limit and also in conformal field theories through the Cardy formula.
Our results confirm this feature for any $q$, beyond the perturbative approach of
\cite{maldacena2016,kitaev2015}. In addition it predicts, also for any $q > 2$, that $\rho(E) \sim \sqrt{E-E_0}$ for $E \to E_0$.
This square root edge, typical of random matrix ensembles has been found in Ref. \cite{erdos2014,cotler2016} but only in the slightly unphysical limit of $q \propto \sqrt{N}$.

 In mesoscopic physics or quantum chaos the occurrence of random matrix theory it is related to full quantum ergodicity in the long time limit \cite{guhr1998}, namely, the system evolves, for sufficiently long times, to a structureless and fully entangled state where only global symmetries characterize the dynamics. 
 These are dynamical features while the spectral density is only related to
 thermodynamical properties which requires further checks  to confirm quantum ergodicity of the SYK model and its gravity dual. For that purpose we have studied
 level statistics in the infrared region where the spectral density is given by Eq.~(\ref{rhocot}).

 \begin{figure}[t!]
 	\centering 
 	\resizebox{0.7\textwidth}{!}{\includegraphics{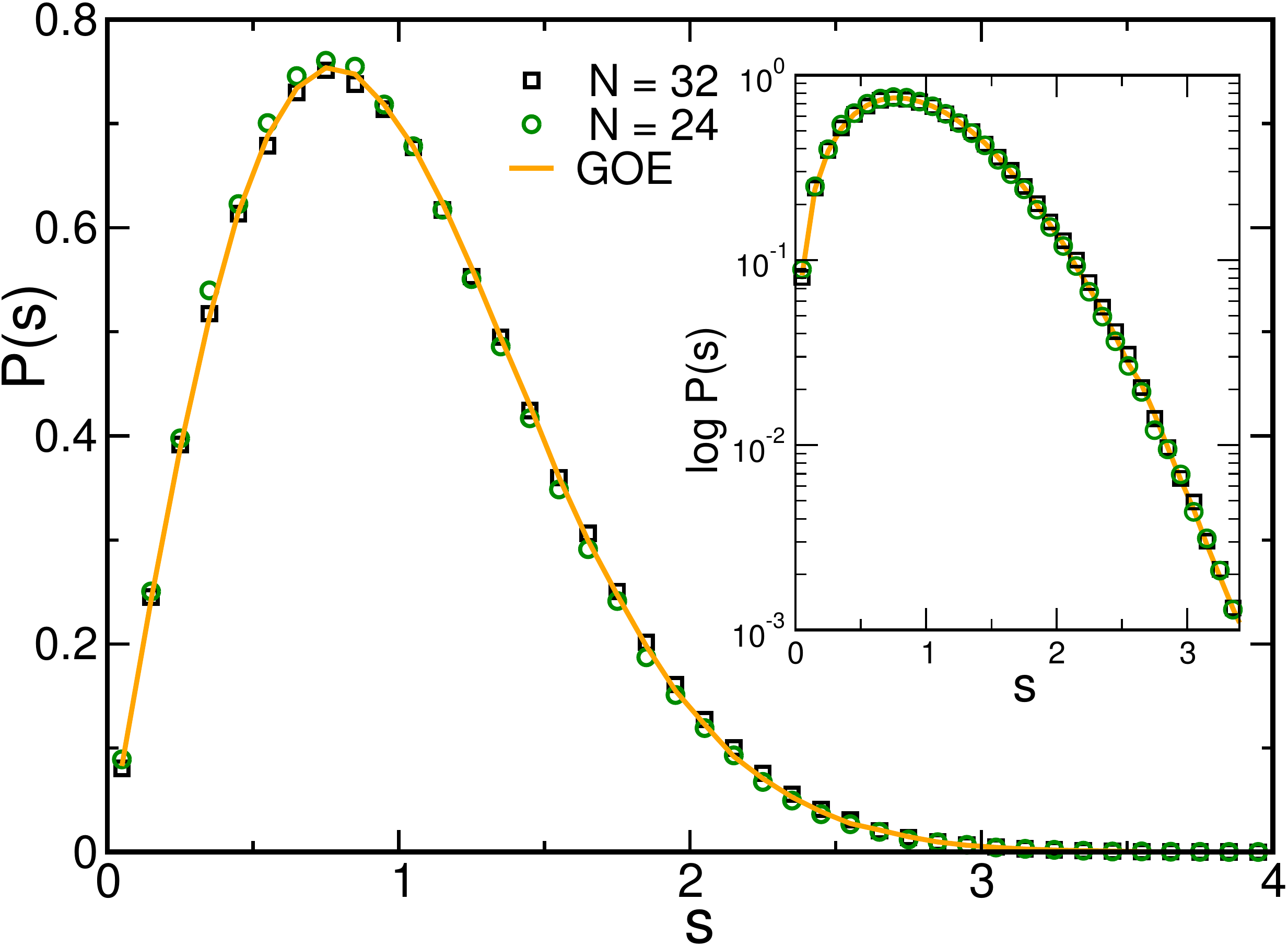}}
      \caption{Level spacing distribution $P(s)$ resulting from exact diagonalization of the SYK Hamiltonian Eq.~(\ref{hami}) for $N = 32$ and $400$ realizations (squares)  and $N = 24$ and $10000$ realizations  (circles). We only consider the infrared part of the spectrum, about $1.5\%$, which is related to the gravity dual of the model. As in the bulk of the spectrum \cite{you2016,garcia2016}, we observe excellent agreement with the Gaussian Orthogonal Ensemble (GOE) result. This strongly suggests that full ergodicity, typical of quantum systems described by random matrix theory, is also a universal feature of quantum black holes.} \label{fig:ps}
\end{figure}
  We note that level statistics of the SYK model  already have been studied  previously  \cite{you2016,garcia2016,cotler2016}. However these papers focus only in the central part of the spectrum that it is not related to properties of the gravity dual. 
 By contrast, we have studied the statistics of the
 low lying eigenvalues, namely, the infrared part of the spectrum.
 Since we are interested in long time dynamics of the order of the Heisenberg time, we investigate the level spacing distribution $P(s)$, defined as the probability to find two neighboring eigenvalues separated by a distance $s=(E_{i+1}-E_i)/\Delta$ where $\Delta$ is the mean level spacing $\Delta$ (see \cite{garcia2016} for details of the calculation like the unfolding procedure).
 In Fig.~\ref{fig:ps} we depict results for $P(s)$ for $N=24$ and $N=32$ considering only $1.5\%$ of the lowest eigenvalues.
 As in the central part of the spectrum \cite{you2016,garcia2016}, it follows closely the prediction of the
 Gaussian Orthogonal Ensemble (GOE).
 The good
 agreement shows that the eigenvalues of the SYK Hamiltonian fluctuate
 according to random matrix theory all the way to the ground state region.
 This shows that the SYK Hamiltonian is chaotic in the infrared domain.
 This is a further confirmation of
 the full ergodicity of the SYK model in the long time limit and, in
 agreement with the result of the previous section, that the distribution
 of the smallest eigenvalue is given by the Tracy-Widom distribution.

 This is a strong indication that not only the SYK model
 but also its gravity dual, a certain type of quantum black hole, are systems whose long time dynamics only depends on global symmetries and always lead to a completely featureless and ergodic quantum state. It is well known that random matrix ensembles are characterized by global symmetries only. It would be interesting to explore whether a similar classification characterizes the long time dynamics of quantum black holes. 

 Nuclear physics is another area in which our results are of potential interest. A central feature of the excitations of complex nuclei is captured by Bethe's \cite{bethe1936} expression that predicts an exponential growth of the density of states for energy close, but not too close, to the edge of the spectrum. Interestingly,
 the exponential growth predicted by the Bethe formula is very similar to that of Eq.~(\ref{rhocot}). Experimental results agree, at least qualitatively with this simple analytical expression. This is not fully understood because interactions are typically strong while Bethe's expression is derived assuming non-interacting fermions in a mean field potential. Our results help explain this puzzle as the exponential growth also occurs in the SYK model, and likely in generalizations thereof, in which fermions are strongly interacting. This is also a strong indication that holography may be a powerful tool to model certain aspects of the physics of strongly interacting nuclei.   

\section{Conclusions}
We have analytically computed the spectral density of the SYK model in the $N~ \gg~1$ limit by an explicit evaluation of the energy moments combined with the use of the Riordan-Touchard formula \cite{riordan1975,touchard1952}. For $N \gg 1$, and $E$ not close to the ground state, it simplifies to 
$\rho(E) \sim \exp [2\arcsin^2(E/E_0) /\log \eta]$.  
In the infrared limit, the analytical expression for the spectral density has a square root singularity, as in random matrix ensembles, followed by an exponential growth. Agreement with exact numerical results is much better than the one expected from the approximations employed in the analytical calculation. 
Our results also agree with the free energy in different limits studied in \cite{maldacena2016} by completely different methods. 

Level statistics in the infrared region are well described by random matrix theory for energy separations of the order of the Heisenberg time. Assuming that even in this deep quantum limit the SYK model still has a gravity dual, our results indicate that, for sufficiently long times, quantum black holes relax universally to a fully ergodic and structureless state where the dynamics is only dependent on the global symmetries of the system. These are exactly the properties of compound
nuclei which have a long history of being described in terms of random matrix theory.

\acknowledgments{A.M.G. thanks Aurelio Berm\'{u}dez and Bruno Loureiro for illuminating discussions. This work  acknowledges partial support from
  EPSRC, grant No. EP/I004637/1 (A.M.G.) and U.S. DOE Grant
  No. DE-FAG-88FR40388 (J.V.).}

%%%%%%%%%%%%%%%%%%%%%%%%%%%%%%%%%%%%%%%%%%%%%%%%%%%%%%%%%%%%%%%%
%\begin{thebibliography}{99}
%\bibliography{library2}
%\end{thebibliography}

\bibliography{library2}

\end{document}